\documentclass{cjaa}
\usepackage{graphicx}
\usepackage{amssymb}
\usepackage{amsfonts}

\begin{document}
\title{Quantum Instability of Magnetized Stellar Objects}

\author{R. Gonz\'{a}lez Felipe
 \inst{1} 
 \and H. J. Mosquera Cuesta\inst{2,3} 
 \and A. P\'{e}rez Mart\'{\i}nez
 \inst{4} 
 \and H. P\'{e}rez Rojas
 \inst{4} 
 }

\institute{Departamento de F\'{\i}sica and Centro de F\'{\i}sica Te\'{o}rica de Part\'{\i}culas,
Instituto Superior T\'{e}cnico, Av.
Rovisco Pais, 1049-001 Lisboa, Portugal\\ 
\and Instituto de Cosmologia, Relatividade e Astrof\'{\i}sica (ICRA-BR), Centro
Brasileiro de Pesquisas F\'\i sicas, Rua Dr. Xavier Sigaud 150,
CEP 22290-180, Urca, Rio de Janeiro, RJ, Brazil\\
\and Centro Latino-Americano de F\'{\i}sica, Avenida Wenceslau Braz
71, CEP 22290-140 fundos, Botafogo, Rio de Janeiro, Brazil\\
\and Instituto de Cibern\'{e}tica, Matem\'{a}tica y F\'{\i}sica,
Calle E No. 309, 10400 La Habana, Cuba\\
 }

\date{}

\abstract{ The equations of state for degenerate electron and neutron gases are
studied in the presence of magnetic fields. After including quantum effects to
study the structural properties of these systems, it is found that some
hypermagnetized stars can be unstable based on the criterium of stability of
pressures. Highly magnetized white dwarfs should collapse producing a supernova
type Ia, whilst superstrongly magnetized neutron stars cannot stand their own
magnetic field and must implode, too. A comparison of our results with a set of
the available observational data of some compact stars is also presented, and
the agreement between this theory and observations is verified.
}

\maketitle

\section{Introduction}

Many stellar objects are known to be endowed with large magnetic fields~(Dryzek
et al.~\cite{dryzek02}, Heyl~\cite{heyl99}, \cite{heyl00}). For instance, white
dwarfs with surface magnetic fields whose strengths range from $10^{5}$ to
$10^{9}$~G have been discovered (see an earlier set of references in Kemp et
al.~\cite{kemp}, Putney~\cite{putney95}, Schmidt \& Smith~\cite{schmidt95},
Reimers~\cite{reimers96}, and the updated list of Suh \&
Mathews~\cite{mathews}). Moreover, magnetic field strengths of the order of
$10^{20}$~G have been suggested to exist in the core of neutron stars or
pulsars. Recently, in Refs. Chaichian et al.~\cite{chaichian} and P\'{e}rez et
al.~\cite{perez}, electron and neutron gases in a strong magnetic field were
considered with the aim to study the equation of state of white dwarf stars,
neutron stars and its relation to supernovae. It was found a pure quantum
effect: the appearance of a {\it ferromagnetic configuration} in the neutron
star interior, which is intrinsically related to the presence of the magnetic
field. Such an effect opens the possibility of a quantum magnetic collapse of
the gas under consideration. This effect is related to the density of the star
and its magnetic field, and as such, it allows one to establish a criterium of
stability taking into account these physical properties.

The presence of a magnetic field drives the loss of the rotational symmetry of
the particle spectrum, which in turn manifests as an anisotropy in the
thermodynamic properties of the system. This behavior can be seen through the
energy-momentum tensor $\mathcal{T}_{\mu\nu}$, if we recall that the external
magnetic field $\mathcal{H}$ induces a magnetization $\mathcal{M}$ in the
medium which is described through the relation $\mathcal{H}=B-4\pi
\mathcal{M}(B)$. Here $B$ is the microscopic magnetic field, which is assumed
to point along the $x_{3}$ axis in what follows. Starting from the
energy-momentum tensor, one can derive the expressions for the pressure
components, longitudinal ($p_{3}$) and perpendicular ($p_{\perp}$) to the
magnetic field,
\begin{equation}
\mathcal{T}_{\perp}=p_{\perp}=-\Omega-B\mathcal{M}\
,\quad\mathcal{T} _{33}=p_{3}=-\Omega\ ,
\end{equation}
where $\Omega$ is the thermodynamic potential.

In classical electrodynamics (Landau \& Lifshitz~\cite{Landau}), i.e., when the
spin interactions are not taken into account, this anisotropy appears but the
results are quite different. Since
\begin{equation}
p_{\perp}=p_{0}+\frac{B^{2}}{4\pi}\ ,\quad
p_{3}=p_{0}-\frac{B^{2}}{4\pi}\ ,
\end{equation}
where $p_{0}$ is the isotropic pressure, one obtains at the classical level
\begin{equation}
p_{\perp}>p_{3}\ .
\end{equation}

This fact explains the oblateness of some astrophysical objects when one
studies them in a classical way. Proper examples of this are provided by
Shapiro \& Teukolsky~\cite{Shapiro} in the case of {\it magnetic white dwarfs},
and by Cardall, Prakash \& Lattimer~\cite{CPL} for the highly magnetized
neutron stars. Contrary to these well-known effects, the interplay of magnetic
and quantum effects imply that some stars get a cigar-like shape along the
$x_3$ axis, and some of them may even collapse.

The purpose of this paper is to discuss in more detail and to exploit some of
the astrophysical consequences of previous works (Chaichian et
al.~\cite{chaichian}, P\'{e}rez et al.~\cite{perez}) regarding the appearance of
hydrodynamic instabilities in strongly magnetized electron and neutron gases as
realizations of the physics taking place inside white dwarfs (WDs) and neutron
stars (NSs), respectively. In particular, these instabilities on the
configuration of any super critically magnetized stellar object are shown to
appear due to the action of quantum-mechanical effects, as the occupation of
the particle  Landau ground state, which is driven by the cooperative particle
spin-magnetic field coupling. These new effects allow us to introduce a new
model to show that some WDs may become ultramagnetized and may collapse in a
kind of SN type Ia, even without reaching the Chandrasekhar mass limit. As a
matter of consistency, we test this theoretical argument in a wide and
extensively studied set of astrophysical sources and show that the conclusions
drawn from our analysis are in complete agreement with present observations of
compact remnant stars, while forbid the existence of some exotic configurations
of them.

The structure of the paper is as follows: In Section~\ref{electrongas} we
present the physics of a magnetized electron gas and apply it to the study of
the stability conditions of white dwarfs. Based on this physics, a new model
for SN type Ia events is presented, which not necessarily depends upon the WD
to reach its critical mass. Section~\ref{neutrongas} considers the case of a
neutron gas, as a model for NSs. We show that the so-called magnetars cannot in
principle be formed based on the instability criterium of anisotropic
pressures, while canonical pulsars would. Finally, in Section~\ref{conclusions}
we present our conclusions.

\section{Magnetized electron gas}
\label{electrongas}

In this section we shall study the thermodynamic properties of the degenerate
electron gas in very intense magnetic fields and density regimes, which for
astrophysical scenarios are the more interesting properties to be investigated.
We shall find the conditions that may lead to the vanishing of the transverse
pressure of the electron gas, as a model of a white dwarf, which may then
undergo a gravitational collapse.

We start by defining the thermodynamic potential as\footnote{Notice that we are
defining the thermodynamic potential per unit volume.}
\begin{equation}
\label{Omega}\Omega= -T \ln\mathcal{Z}\ ,
\end{equation}
where $T$ is the gas temperature and $\mathcal{Z}$ the partition function. In
the case of a magnetized degenerate electron gas, the sum over Landau levels
appear in the electron gas configuration and, therefore, the thermodynamical
potential can be expressed as
\begin{equation}
\Omega_{e} =-\Omega_{0}\sum_{0}^{n_{\mu}}a_{n}B \left[
x\sqrt{x^{2}-1-2nB/B_{c}} -(1+2nB/B_{c}) \times
\ln\left(\frac{x+\sqrt{x^{2}-1-2nB/B_{c}}}
{\sqrt{1+2nB/B_{c}}}\right)\right] \ , \label{Omegae}
\end{equation}
where we define $\Omega_{0}= em^2/(4\pi^{2}\hbar^2 c^2)$,
$a_{n}=2-\delta_{0n}$, $x=\mu_{e}/m_{e}$, $B_{c}=m_{e}^2/(e\hbar
c)$, with $\mu_{e}$ as the chemical potential for electrons and
$n_{\mu}$ corresponds to the maximum Landau level for a given
Fermi energy and magnetic field strength. The maximum occupancy of
these Landau levels is defined as
\begin{equation}
n_{\mu}=I\left[\frac{B_{c}}{2B}\left(x^{2}-1\right)\right]  \ , \label{nmu}
\end{equation}
$I[x]$ denotes the integer part of $x$.

The mean density of particles  is given by $N_{e} = - \partial
\Omega_{e}/\partial\mu_{e} $. In the degenerate limit, i.e., at zero
temperature, one obtains
\begin{equation}
N_{e} = N_{0} \left(\frac{B}{B_{c}}\right) \sum_{0}^{n_{\mu}} a_{n}
\sqrt{x^2-1-2nB/B_{c}}\ ,
\end{equation}
where $N_{0} =  m_{e}^3/(4 \pi^{2}\hbar^3 c^3)$. Bearing in mind that the
magnetization is defined as $\mathcal{M}_{e}=-\partial\Omega_{e}/\partial B\,$,
from Eq.~(\ref{Omegae}) one obtains
\begin{equation}
 \mathcal{M}_{e}
=\mathcal{M}_{0} \sum_{0}^{n_{\mu}}a_{n}\left[x
\sqrt{x^{2}-1-2nB/B_{c}} -(x^{2}+4nB/B_{c})
 \times \ln\left(\frac{x+\sqrt{x^{2}-1-2nB/B_{c}}}
{\sqrt{1+2nB/B_{c}}}\right)\right]  \label{termo} \,,
\end{equation}
where we define ${\mathcal M_{0}}=e m_{e}^2/(4\pi^{2}\hbar^2 c^2)$. We stress
that the magnetization, as well as other thermodynamic quantities, is a
function of $B$ and $N_{e}$. Hence, one can write the expression of
$\Omega_{e}$ in terms of the magnetization as
\begin{equation}
\Omega_{e} = -B \mathcal{M}_{e}-\Omega_{0}\sum _{n=1}^{n_{\mu}} n\,
\ln\left(\frac{x + \sqrt{x^{2}-1-2nB/B_{c}}}{\sqrt{1+2nB/B_{c}}}\right)\,.
\label{omeMag}
\end{equation}

Since the thermodynamic potential is a measure of the internal pressure in the
gas, then it becomes especially interesting to discuss the equation of state of
the system at this point. It is worth to note that the latter receives
contributions from the partial pressures of the several species involved. By
calculating the energy-momentum tensor of this gas (made up of electrically
charged particles) one obtains different equations of state for the directions
parallel, $p_{3}$, and perpendicular, $p_{\perp}$, to the magnetic field $B$.
They read
\begin{equation}
p_{3} = -\Omega_{e}\ ,
\end{equation}
and
\begin{equation}
\label{pperfinal}p_{\perp}=\frac{2 e^{2} B^{2}}{\pi^{2}(\hbar c)}
\sum_{n=0}^{n_{\mu}} n\, \ln\left(\frac{x + \sqrt{x^{2} -
1-2nB/B_{c}}}{\sqrt{ 1+2nB/B_{c}}}\right)\ .
\end{equation}

It is readily verified from these expressions that for a positive magnetization
the transversal pressure exerted by the charged particles in the presence of a
magnetic field  is smaller than the longitudinal one by the amount $B
\mathcal{M}_{e}$.

There are some situations where densities and magnetic fields are such that
only the first Landau level of the particle spectrum is occupied.
Eq.~(\ref{pperfinal}) shows that when the electrons are confined to the Landau
ground state, one can write
\begin{equation}
p_{\perp}=-\Omega_{e}-B\mathcal{M}_{e} = 0\ , \label{p-perpendicular}
\end{equation}
while higher excited Landau states provide a positive contribution to the
pressure. If the dominant contribution to the pressure comes from the electron
gas, then the vanishing of $p_{\perp} $ means that the pressure due to the
gravitational force, of order $G M^{2}/R^{4}$ ($R$ is the radius of the star),
cannot be compensated, and thus there appears an instability which leads to an
abrupt implosion of the electron gas. It is interesting therefore to find the
conditions for the occurrence of this confinement to the ground state ($n = 0$)
and, consequently, for the triggering of the gravitational collapse of the
star.

\begin{figure}
[ptb]
\begin{center}
\includegraphics[width=8.5cm] {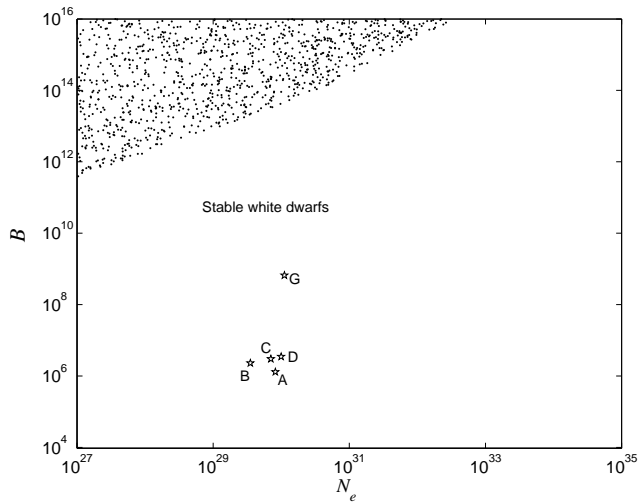} \caption{The instability
region (dotted area) in the $(N_{e},B)$-plane for a magnetized electron gas. A
WD star whose configuration lies inside the dotted region collapses due to the
vanishing of the transverse pressure
$p_{\perp}=-\Omega_{e}-B\mathcal{M}_{e}\,$. The labelled points represented by
stars correspond to some of the stable white dwarf configurations given in
Table~\ref{tab1}.} \label{fig1}
\end{center}
\end{figure}

In general, from Eq.~(\ref{nmu}) one can write
\begin{equation}
n_{\mu} \simeq 2 \times 10^{-21}\frac{N_{e}^{2}}{B^{3}}\,,
\end{equation}
which essentially defines an instability criterium for all magnetized  stars in
the plane ($N_{e},B)$. In  Fig.~\ref{fig1} we show the instability region
(dotted area) for a magnetized electron gas in that plane. Any star for which
its structural configuration places it inside the dotted region  should undergo
a transverse collapse due to the vanishing of the pressure transverse
(perpendicular) to the magnetic field.

\begin{table}[ptb] \centering\renewcommand{\tabcolsep}{0.8pc}

\begin{tabular}
[c]{cccc}\hline\hline WD & $M/M_{\odot}$ & $N_{e}/N_{0}$ & $B$ (in
Gauss)\\\hline
A & 1.28 & $1.4$ & $1.3\times10^{6}$\\
B & 0.9 & $0.6$ & $2.3\times10^{6}$\\
C & 1.2 & $1.2$ & $3\times10^{6}$\\
D & 1.31 & $1.7$ & $3.5\times10^{6}$\\
E & 0.76 - 1.0 & $0.3-0.9$ & $>3\times10^{8}$\\
F & $>1.0$ & $0.9$ & $3.2\times10^{8}$\\
G & 1.35 & $1.9$ & $6.6\times10^{8}$\\
H & ? & ? & $\simeq 10^{9}$\\
I & ? & ? & $\gtrsim10^{9}$\\\hline\hline
\end{tabular}
\caption{Mass $M$, electron number density $N_{e}$ and surface magnetic field
$B$ for some typical magnetic white dwarfs: A = PG 0136 + 251, B = PG 2329 +
267, C = 1RXS J0823.6-2525, D = PG 1658 + 441, E = LB 11146B, F = Grw +
70$^{\circ}$8247, G = RE J0317-858, H = PG 1031 + 234 , I = GD 229. The above
data were taken from Suh \& Mathews~\cite{mathews}. Here
$N_{0}=m_{e}^{3}/(4\pi^{2}\hbar^3 c^3)\simeq5.9\times10^{29}$~cm$^{-3}$ and
$M_{\odot}\simeq1.99\times10^{33}$~g is the solar mass.}\label{tab1}
\end{table}

\subsection{White Dwarfs}

Recent observations have pointed out that supernovae type Ia (SNIa), considered
as standard candles in the redshift range $0.1\leq z\leq1$, seem to provide
evidence for our Universe to be accelerating (Riess et al.~\cite{riess98},
Perlmutter et al.~\cite{perlmutter99}), driven by some kind of dark energy: a
cosmological constant or quintessence field. However, the premise underlying
this suggestion needs to be firmly established on astrophysical foundations. In
order to settle out this issue in a conclusive way, the redshift independent
nature of SNIa should be demonstrated (Karl et al.~\cite{karl03}, Napiwotzki et
al.~\cite{napi03}, Hillebrandt et al.~\cite{niemeyer02}), since only a complete
understanding of those events can give the required confidence, if any, to the
idea that they are indeed standard candles. In this paper we introduce a new
scenario for the SN type Ia events based on the physics presented above, which
indicates that another kind of collapse may produce SNIa outbursts.

The leading scenario for SNIa suggests that these explosions stem from the
complete disruption of a white dwarf (WD) induced by accretion of mass from a
companion star, which forces it to surpass the WD Chandrasekhar mass limit
$M_{\rm crit}\sim 1.445\ M_{\odot}$ (Livio~\cite{livio2000}). The abrupt
conversion of nearly $1~M_{\odot}$ of C/O to $^{56}$Ni and the $^{56}$Ni
subsequent decay to $^{56}$Fe releases, on the deserved timescale, an amount of
energy just as large as needed to power the observed luminosity and kinematics
of a SNIa. In addition, this scenario provides a consistent explanation for the
absence of hydrogen in these events. Although this model is well motivated, it
has been shown to contain several drawbacks that tend to disfavor it
(Hillebrandt et al.~\cite{niemeyer02}).

In an alternative view, recent Hubble Space Telescope observations suggest that
a good candidate to SN type Ia progenitors could be coalescing binary WD-WD
dubbed as {\it blue strugglers}. These objects are a class of stars that live
in globular clusters orbiting in the outskirts of spiral galaxies. A particular
set of these candidates to SNIa are mergers of unequal mass WD-WD binaries that
come together because of angular momentum loss via gravitational radiation (GR)
emission (Karl et al.~\cite{karl03}, Napiwotzki et al.~\cite{napi03}). The
timescale ($\sim10^{6}$ yr) for the system to coalesce is dictated by GR, so
that the donor companion has the right time scale to fill its Roche lobe
driving then mass transfer onto its heavier orbital partner
(Nomoto~\cite{nomoto1998}). However, there exists a debate in the literature
whether an explosion does occur under such a dynamics. Some authors contend
that view and suggest that more likely a \textquotedblleft
silent\textquotedblright\ accretion-induced collapse (AIC) to form a remnant
neutron star must take place (Mochkovitch et al.~\cite{mochkovitch1997}). In
overall, the issue seems contentious.

\subsection{Quantum-Magnetic Collapse of a WD as Model of SNIa}

In the lines of the theory presented earlier (P\'{e}rez et al.~\cite{perez}), we
suggest here a quite different mechanism in which, for the case of young,
rapidly rotating, highly magnetized and massive C/O WDs, a direct collapse of a
single WD should indeed occur. The overall process is driven by both the
quantum and magnetic effects described above.

Rapidly rotating white dwarfs have been studied theoretically (see Shapiro \&
Teukolsky~\cite{Shapiro} and the extended list of references therein). Despite
earlier thoughts dismissing that possibility, based on the fact that the
by-the-time observed H lines in most WD spectra exhibited narrow cores, recent
observations, however, confirm that such objects do exist (Karl et
al.~\cite{karl01}) among the more than 2400 WDs cataloged up today. The
shortest rotation period that has been clearly measured is the 33~s of AE Aqr,
while a good candidate for a slightly faster spin is WZ Sge with a period of
28~s. Nonetheless, there appears to exist a faster class of WDs spinning with
periods of $P_{crit} \sim 27$~s, dubbed as DQ Herculis, which, on the other
hand, are magnetized WD accreting matter in binaries christened cataclysmic
variables.

For non-magnetic white dwarfs, meanwhile, the theoretical view suggests that
there exists a critical rotation period (or frequency) of uniformly spinning
WDs for which mass shedding at the star equator takes place. The instability
sets in for a Keplerian rotation frequency
\begin{equation}
 \Omega_K = 0.67 (\pi G \bar{\rho})^{1/2}\, ,
\end{equation}
where $\bar{\rho}$ is the WD mean density. That relation implies a rotation
period $P_K \sim 17$~s, for stars with mean density in the range $10^4\,{\rm
g/cm}^3 \leq \rho \leq 10^7\,{\rm g/cm}^3$. However, for WD stars rotating
differentially at the relevant temperature the threshold period can be much
smaller than the above limit. This is a crucial issue that we will take
advantage of in developing our model. According to Shapiro \&
Teukolsky~\cite{Shapiro}, in a differentially rotating WD the ratio between the
frequency at the center and at its surface is
\begin{equation}
 \frac{\Omega_{equator}}{\Omega_{center}} \leq \frac{1}{5}\,,
\end{equation}
for almost all the equilibrium configurations with masses in the range
$0.4-0.9\, M_\odot$. Massive WD models above $1.4\,M_\odot$ possess surface
velocities around $3000-7000$~km/s, which suggests periods around 1~s.

A WD in a binary system may accrete material from its orbital partner in such a
way that its angular momentum can rise significantly, and it may rotate
differentially. As the material accreted transfers energy and angular momentum,
the primary WD star will be spun-up over the time scale
\begin{eqnarray}
\Delta T^{s}_{u} & \simeq & \frac{J}{\dot{J}} = \frac{2 M R^2
\Omega}{5 (GMR)^{1/2}} \frac{1}{\dot{M}} = 4 \times 10^6\,{\rm
yr.} \left(\frac{M}{{M_\odot}}\right)^{1/2}  \;  \nonumber \\
& \times & \left(\frac{R}{10^{-2} R_\odot}\right)^{3/2}
\left(\frac{1\, \rm s}{P}\right) \left(\frac{10^{-8} M_\odot \rm
y^{-1}}{\dot{M}}\right)\; ,
\end{eqnarray}
where $J$ is the angular momentum accreted from the companion, $\dot{J}$ - the
applied torque, $P$ - the rotation period, $\dot{M}$ - the  accretion  rate,
and $R$ is the WD radius in units of the solar radius $R_\odot$. Notice that a
higher accretion rate, as in unequal WD-WD binaries, may reduce the spin-up
time scale. Therefore, an essential ingredient of this model is the existence
of a WD primary in a binary system which is old enough to having had time to
accrete up to the point of being spun-up to its break-up velocity at the
equator.

Once we have the star differentially rotating near its break-up period, the
conditions are reached for the $\alpha-\Omega$ dynamo mechanism (Duncan \&
Thompson~\cite{duncan}, Thompson \&
Duncan~\cite{thompson93,thompson95,thompson96}) to start to amplify the WD
initial magnetic field. The maximum magnetic field achievable through this
process can be as strong as (the equipartition of pressure and kinetic energy)
\begin{equation}
\frac{B^2}{8 \pi} = \frac{M_{\rm WD} \Omega_{\rm WD}^2 R^2_{\rm WD} }{(8 \pi/3)
R^3_{\rm WD}} \longrightarrow B^s_{c} \sim 3\times10^{14} \left(\frac{P_{i}}{1\
\mathrm{s}}\right)^{-1}\,{\rm G} \; ,
\end{equation}
with $B^s_{c}$ being the supercritical $B$-field and $P_{i}$ the shortest
period reached upon undergoing the $\alpha-\Omega$ dynamo phase. This $B^s_{c}$
should be generated as the differential rotation is smoothed by growing
magnetic stresses inside the WD star. For the typical WD mass, radius and
rotation frequency given above, the associated spin energy in the process reads
\begin{equation}
E_{\Omega i} = \frac{1}{2} M_{\rm WD} \Omega_{\rm WD}^2 R^2_{\rm WD} \sim
10^{51} \left(\frac{P_{i}}{1\ \mathrm{s}}\right)^{-2}\,{\rm erg} \; ,
\end{equation} which is similar to the characteristic energy of most SNIa
events. This energy plus the gravitational binding energy, which is of the same
order of magnitude as the above, would be released during the implosion of the
star. The computed amplified magnetic field\footnote{Notice that the final
$B$-field is almost independent of the WD initial $B$-field.} plus the matter
density of a typical WD will put the star in the instability region showed in
Fig~\ref{fig1}, according to Eq.~(\ref{p-perpendicular}), where it must undergo
a gravitational collapse. Thus, the higher the WD initial $B$-field, the
shorter the break-up spin-up time scale. This process may take place well in
advance the star reaches the Chandrasekhar critical mass of $1.445\, M_\odot$.

Since the time scale estimated above is shorter than the age of our universe,
certainly old WD stars may accrete enough material so as to rotate at its
break-up velocity. At that point its initial $B$-field might be enlarged via
the dynamo mechanism, which in turn may drive the star to explode as a
consequence of the quantum-magnetic collapse it undergoes. Those systems may be
triggering the observed SNIa events.

Whenever the physical properties of the WD star (see Table~\ref{tab1}) places
it above the instability region depicted in Fig.~\ref{fig1}, the WD must
implode triggering a supernova-like event with the characteristics of a SNIa.
The WD dynamical timescale and the abrupt combustion of its constituent
material (electron capture into protons, and rapid process nucleosynthesis to
$^{56}$Ni) is to produce the energetics as required on the observational basis.
The main reason for this claim is that both characteristics, i.e. the WD rapid
spinning and age, can amplify (via the $\alpha-\Omega$ dynamo mechanism) the
remnant magnetic field ($\sim 10^{8}$~G) up to a critical value to which it may
be unstable and thus collapses. In fact, there is strong evidence for young and
highly magnetized WDs, as for instance the object RE J0317-858 listed in
Table~\ref{tab1}. The process we describe is alike to the accretion induced
collapse quoted above, but originated from the cooperative action of two
different physical effects. At the end, the SNIa explosion must be accompanied
by a gamma-ray burst of energy $\sim10^{51}$~erg. This is an interesting
signature to look for. A further testable prediction of this picture could be
to find no remnant at all after a SNIa! Such a result could be interpreted as
if the WD had collapsed into a black hole or a black string, driven by the
combination of the quantum and magnetic effects discussed here, and purported
in P\'{e}rez et al.~\cite{perez}.

Were this model, i.e. the collapse of heavy and strongly magnetized WD,
realized in nature, it would significantly help to clarify whether SNIa could
indeed be considered as legitimate standard candles, since for the collapse to
take place in this case the WD mass must be in the range $0.9-1.35~M_\odot$,
for which the total binding energy released is practically the same $\sim
10^{51}$~erg. Of course, some other observational constraints play some r\^ole,
as for instance the statistics of massive WD-WD binaries, which is relatively
small (Karl et al.~\cite{karl03}, Napiwotzki et al.~\cite{napi03}).

In brief, several white dwarfs endowed with strong magnetic fields have been
observed and studied (Kemp et al.~\cite{kemp}, Suh \& Mathews~\cite{mathews},
see Table~\ref{tab1}). From the analysis of the observational data (Barstow et
al.~\cite{barstow}, Greenstein et al.~\cite{greenstein}) one can see that our
model is in close agreement with those observations. These stars are stable
objects. On the other hand, if those WD listed in Table~\ref{tab1} were given
an amplified supercritical ($B^s_u$) magnetic field, then they must implode.

\section{Magnetized neutron gas}
\label{neutrongas}

Most of the observed neutron stars are pulsars, i.e., fast rotating neutron
stars with strong magnetic fields (Lorimer~\cite{lorimer99}). They consist
mainly of neutron matter with a high central density. As in the case of WDs, a
similar study can be performed for neutron stars which are properly described
by a degenerate neutron gas (Shapiro \& Teukolsky~\cite{Shapiro}). In this
case, it is straightforward to calculate the thermodynamic quantities starting
from the neutron density $N_{n}$. Assuming that the uniform magnetic field $B$
lies along the $x_{3}$-axis, and using the Dirac equation for neutral particles
with anomalous magnetic moment (Bagrov \& Gitman~\cite{Bagrov1990}) propagating
through this field, we get the spectrum as
\begin{equation}
E_{n}(p,B,\eta)=\sqrt{p_{3}^{2}+\left( \sqrt{p_{\perp}^{2} +
m_{n}^{2}} + \eta|q_{n}|B\right) ^{2}}\ , \label{En}%
\end{equation}
where $p_{3}, p_{\perp}$ are the momentum components along and perpendicular to
$B$, respectively; $m_{n}$ is the neutron mass and $q_{n}\simeq-1.91\mu_{N}$ is
the neutron magnetic moment ($\mu_{N}=e/(2m_{p})$ is the nuclear magneton), and
$\eta=\pm1$ are the $\sigma_{3}$ eigenvalues corresponding to the two possible
orientations (parallel or antiparallel) of the neutron magnetic moment with
respect to the magnetic field.

The thermodynamic properties of a degenerate neutron gas are easily obtained
following the same procedure as in Section II and it was the scope of Ref.
P\'{e}rez et al.~\cite{perez}, which we revisit now to discuss their astrophysical
implications.

Firstly, the neutron density $N_{n}$, thermodynamical potential $\Omega_{n}$,
and magnetization $\mathcal{M}_{n}$ read, respectively,
\begin{eqnarray}
N_{n}&=&N_{0}\sum_{\eta=1,-1}\left[
\frac{f_{\eta}^{3}}{3}+\frac{\eta y(1+\eta y)f_{\eta}}{2}-\frac{\eta
yx^{2}}{2}s_{\eta}\right] \
,\label{Nn}\\
\Omega_{n}&=&-\Omega_{0}\sum_{\eta=1,-1}\left[ \frac{xf_{\eta}^{3}}
{12}+\frac{(1+\eta y)(5\eta y-3)xf_{\eta}}{24} + \frac{(1+\eta
y)^{3}(3-\eta y)}{24}L_{\eta}-\frac{\eta yx^{3}}
{6}s_{\eta}\right] \ ,\label{Omegan}\\
\mathcal{M}_{n}&=&-\mathcal{M}_{0}\sum_{\eta=1,-1}\eta\left[
\frac{(1-2\eta y)xf_{\eta}}{6} -\frac{(1+\eta y)^{2}(1-\eta
y/2)}{3}L_{\eta}+\frac{x^{3}} {6}s_{\eta}\right] \ . \label{Mn}
\end{eqnarray}
Here we define the numerical constants:
$N_{0}=m_{n}^{3}/(4\pi^{2}\hbar^3 c^3) \simeq 2.73\times 10^{39}$
cm$^{-3}$, $\Omega_{0} =N_{0}m_{n}\simeq 4.11\times 10^{36}$ erg
cm$^{-3}$ and $\mathcal{M}_{0}=N_{0}q_{n} \simeq 2.63\times
10^{16}\, {\rm erg/(G\,cm}^3)$. We also set
\begin{equation}
x=\frac{\mu_{n}}{m_{n}}\ ,\quad y=\frac{B}{B_{n}}\ ,
\end{equation}
with $B_{n}=m_{n}/q_{n}\simeq 1.56\times 10^{20}$~G, and introduced the
notations:
\begin{equation}
f_{\eta} = \sqrt{x^{2}-(1+\eta y)^{2}},\quad s_{\eta}=\frac{\pi}{2}
-\arcsin\left( \frac{1+\eta y}{x}\right) ,\quad L_{\eta} =\ln\left(
\frac{x+f_{\eta}}{1+\eta y}\right) \ .
\end{equation}

In the limit $B = 0$, Eqs.~(\ref{Nn}) and (\ref{Omegan}) reproduce the usual
density and thermodynamic potential of a relativistic Fermi gas at zero
temperature (Fradkin~\cite{Fradkin}). We see that the magnetization
$\mathcal{M}_{n}$ is a nonlinear function of $B$, and since $\mathcal{M}_{n}
\geq0$ the magnetic response of the neutron gas is {\it ferromagnetic} (P\'{e}rez
et al.~\cite{perez}).

As we mentioned in the introduction, the key ingredient brought by the
anisotropy of pressures is precisely that the magnetization is positive. In
fact, one finds that the condition $p_{3}> p_{\perp }$ holds for any study of a
highly magnetized Fermi gas. In particular, we remark that for a neutron gas
with anomalous magnetic moment this condition is fulfilled.

We also notice that if $x=1+\eta y$, then $f_{\eta}=s_{\eta} = L_{\eta}=0$. For
$x < 1+y$ the thermodynamic potential $\Omega_{n}$, and consequently, all the
thermodynamic quantities in the system become complex. The condition $x=1+y$
defines a curve in the $(N_{n},B)$-plane described by the relation
\begin{equation}\label{Nncurve1}
N_{n}=N_{0}\left\{ y^{5/2}+\frac{5}{3}y^{3/2} +
\frac{1}{2}y(1+y)^{2}\left[ \frac{\pi}{2} - \arcsin\left(
\frac{1-y}{1+y}\right) \right] \right\} \ ,
\end{equation}
which delimits the region where the pressure becomes complex. Since $y\ll1$ for
the magnetic field strengths under consideration ($B_s \sim 10^{15}~$G in the
NS surface), we can expand Eq.~(\ref{Nncurve1}) around $y=0$ to obtain the
approximate expression
\begin{equation}
N_{n}\simeq\frac{8}{3}N_{0}\ y^{3/2}.
\end{equation}

It is also straightforward to find the curve which corresponds to the vanishing
of $p_{\perp}$. Writing $x=1+(1+a)y$, with $ a>0$ in
Eqs.~(\ref{Nn})-(\ref{Mn}), and expanding around $y=0$ we obtain to leading
order in $y$ :
\begin{equation}
N_{n}\simeq\frac{2\sqrt{2}}{3}N_{0}\ y^{3/2}\left[
(a+2)^{3/2}+a^{3/2} \right] \ , \label{Nnap}
\end{equation}

\begin{equation}
p_{\perp}
\simeq\frac{2\sqrt{2}}{15}\Omega_{0}\frac{y^{5/2}}{\sqrt{2+a}
}\left[ 2a^{3}+7a^{2}+4a+a^{3/2}(5+2a)\sqrt{2+a}-4\right] \ .
\label{pperap}
\end{equation}

The solution of the equation $p_{\perp}=0$ is given by $a=a_{0}\equiv
\frac{3}{5}\sqrt{5}-1\simeq0.34 $. Substituting this value into
Eq.~(\ref{Nnap}) we find
\begin{equation}
N_{n} \simeq\frac{2\sqrt{2}}{3}N_{0}\ y^{3/2}\left[ \left( \frac{3}
{5}\sqrt{5}+1\right) ^{3/2}+\left( \frac{3}{5}\sqrt{5}-1\right)
^{3/2}\right] \simeq3.57N_{0}\ y^{3/2}\; . \label{Nncurve2}
\end{equation}

In Table II we present some results for neutron stars obtained by using other
equations of state (EOS), and taking into account that $B$-fields in the core
may fulfill the dipolar law, i.e., that $B$ obeys the relation $B =
B_{s}(R_{s}^3/r^3)$, at least for the dipole component; here $R_{s}$ is the
star (surface) radius and $r$ is any inner radius of the star. Our results were
obtained by using the data from Cardall, Prakash \& Lattimer~\cite{CPL}, where
the authors performed ``realistic'' numerical simulations of strongly
magnetized neutron stars. We identify $B_{s} = B_{pole}$, as of usage in
Cardall, Prakash \& Lattimer~\cite{CPL}. Under these conditions, and based on
the proposed criterium of stability, we can conclude that those neutron stars
would be unstable objects.

\begin{table}[ptb] \centering\renewcommand{\tabcolsep}{0.85pc}%
\begin{tabular}
[c]{ccccc}\hline\hline NS & EOS & $M/M_{\odot}$ & $N_{n}/N_{0}$
&$B$ (in Gauss)\\\hline
A & Pol2 & 2.31 & $0.019$ & $5.1\times10^{20}$ \\
B & Pol2 & 2.37 & $0.11$ & $1.48\times10^{20}$\\
C & BJI & 1.96 & $0.18$ & $9.1\times10^{20}$\\
D & BJI & 2.12 & $0.25$ & $2.86\times10^{21}$\\
E & PandN & 1.72 & $0.26$ & $5.8\times10^{20}$\\
F & PandN & 1.86 & $0.35$ & $1.6\times10^{21}$\\
G & Akmal & 2.22 & $0.21$ & $6.4\times10^{20}$\\
H & Akmal & 2.53 & $0.19$ & $3.2\times10^{21}$\\
I & PCL & 2.09 & $0.054$ & $8.7\times10^{20}$\\
J & PCL & 1.91 & $0.25$ & $1.6\times10^{21}$\\
K & PCLhyp & 2.02 & $0.30$ & $6.9\times10^{20}$\\
L & PCLhyp & 1.76 & $0.04$ & $1.38\times10^{21}$\\\hline\hline
\end{tabular}
\caption{Mass $M$, neutron number density $N_n$ and magnetic field $B$ for some
neutron star equilibrium configurations computed in Ref.~\cite{CPL} using
different equations of state (EOS). Here $N_{0}= m_n^3/(4\pi^2 \hbar^3 c^3)
\simeq 2.73 \times 10^{39}\,\rm cm^{-3}$.} \label{tab2}
\end{table}

\begin{figure}[ptb]
\begin{center}
\includegraphics[width=8.5cm ] {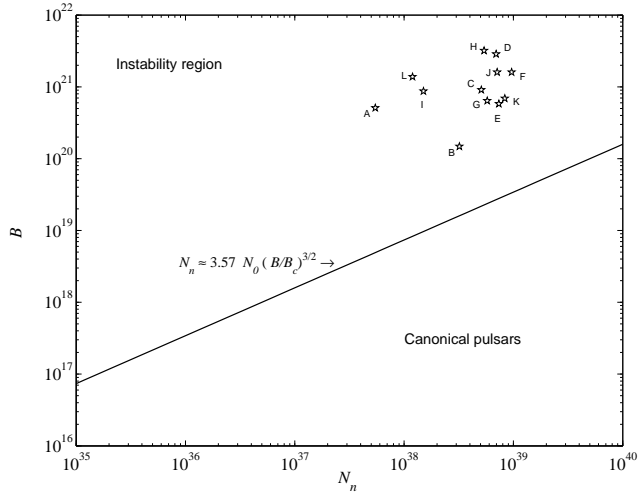} \caption{The instability
region in the $(N_{n},B)$-plane for a magnetized neutron gas, as a model of a
pulsar. A star whose configuration lies above the solid curve defined by
Eq.~(\ref{Nncurve2}) should collapse due to the vanishing of the transverse
pressure $p_{\perp}=-\Omega_{n}-B\mathcal{M}_{n}\,$. The labelled points
represented by stars correspond to the neutron star configurations listed in
Table~\ref{tab2} and computed using different EOS. The core of any star being
trapped in the instability region must implode or collapse as in the quark-nova
model (Ouyed et al.~\cite{ouyed02}).} \label{fig2}
\end{center}
\end{figure}

In Fig.~\ref{fig2} we plot the curve defined by Eq.~(\ref{Nncurve2}) in the
$(N_{n},B)$-plane. For values of $x>1+(1+a_{0})y$ the gas pressure is positive
and the neutron gas is stable. One can verify, on the other hand, that if
$x\leq1+(1+a_{0})y$ the pressure becomes negative or vanishes, thus leading to
the transverse collapse of the star.

\subsection{Magnetars}

It has been recently suggested that there might be stellar objects, known as
soft gamma repeaters (SGR), whose magnetic field inferred from the spindown of
their believed associated pulsars\footnote{See also Cuesta~\cite{herman2000}
for a different approach.} seems to exceed the critical value:
$B_{c}=m_{e}^{2}/(e \hbar c) \simeq 4.41 \times 10^{13}$~G. In particular, it
is claimed that these are newly born neutron stars (Mazets et
al.~\cite{mazets}, Murakami et al.~\cite{murakami}, Kouveliotou et
al.~\cite{kouve}), also called magnetars (Duncan \& Thompson~\cite{duncan},
Thompson \& Duncan~\cite{thompson93,thompson95,thompson96}), with surface
magnetic fields as strong as $10^{15}$~G. According to Duncan \&
Thompson~\cite{duncan}, Thompson \&
Duncan~\cite{thompson93,thompson95,thompson96}, a neutron star with period
$P_{i}\sim1$~ms can support an efficient $\alpha-\Omega$ dynamo. Since neutron
stars are formed with significant differential rotation, the associated energy
is $E_{\Omega i} \sim10^{52}(P_{i}/1\ \mathrm{ms})^{-2}$~erg. Thus magnetic
fields as strong as $3\times10^{17}(P_{i}/1\ \mathrm{ms})^{-1}$~G can be
generated as the differential rotation is smoothed by growing magnetic stresses
inside the star. After the available energy is released in the outermost parts
of the star, vigorous convection continues to generate much stronger magnetic
fields than any previous phase of stellar convection, producing fields larger
than the one required to balance the gravitational binding energy density
(Duncan \& Thompson~\cite{duncan}, Thompson \&
Duncan~\cite{thompson93,thompson95,thompson96}). Thus their result suggests
that the $\alpha-\Omega$ dynamo operating in 1 ms neutron stars might generate
surface dipole fields largely stronger than $10^{13}$~G.

For this class of compact magnetized stars with $B_s \sim 3\times10^{17}
(P_{i}/1\ \mathrm{ms)^{-1}}$~G, our description above predicts that such
objects should implode to a magnetically ($B\leq B_{c}$) stable configuration,
which is not a magnetar! This result, previously obtained in P\'{e}rez et
al.~\cite{perez}, was later confirmed by Khalilov
(Khalilov~\cite{Khalilov:2002rz}) and Chakrabarty's group in a very interesting
series of papers (Ghosh \& Chakrabarty~\cite{chakrabarty}, Ghosh et
al.~\cite{Ghosh:2002dt}, Mandal \&
Chakrabarty~\cite{Mandal:2002eq,Mandal:2002sw,Mandal:2003th}). In particular,
Khalilov in his stability analysis of a degenerate neutron gas in chemical
equilibrium with a background of electrons and protons, also included the
contribution from the anomalous magnetic moment of the fermions composing the
star, as done in Ref. P\'{e}rez et al.~\cite{perez}. We notice, however, that in
Ref. Khalilov~\cite{Khalilov:2002rz} a longitudinal collapse is found instead
of a transversal one. A detailed discussion of the approach followed in the
latter work, which seems to us inappropriate, will be presented elsewhere
(P\'{e}rez et al.~\cite{perez05}).

As commented above (cf. Fig.~\ref{fig2}) some neutron stars described by other
EOS producing $B_s \sim 3 \times 10^{17}$~G fall in the region of instability
defined by our theory. The attentive reader should note that this magnetic
field strength is identical to the one invoked by Kondratyev
(Kondratyev~\cite{kondratyev}) in an attempt to explain the statistical
properties of magnetars. By relating intervals of intense activity with sharp,
step-like changes of magnetization, due to structural inhomogeneities in the
crust and release of its stored magnetic energy, Kondratyev's letter claims
that this could be the mechanism triggering SGRs activity. The demagnetization
takes place as random jumps associated with magnetic avalanches and sharp
energy injection into the magnetosphere develop. Because the nucleons populate
discrete energy levels in such a matter, for field strengths $B \sim
10^{16-17}$~G (and density near $\rho_{\mathrm{nuclear}}$), where the energy
levels crossover, the nuclei structure changes abruptly to a stepwise field
dependence of the nucleus magnetic moment on the nucleon magneton. According to
this author, the decay of this configuration would then drive the SGRs bursting
phases.

However, for a crustal {\it multipolar} field configuration one can expect, at
the stellar interior, a field strength much higher than $B \sim 10^{17}$~G, at
least for the dipole component. This is an idea the astrophysical community is
well acquainted with. If this is the case, then our analysis jumps into scene
showing (see Fig.~\ref{fig2}) that much earlier in the neutron star life it
should have collapsed. Notice further that Woods et al.~\cite{Woods:2001ks},
after studying large torque variations in two SGRs, concluded that within the
context of the magnetar model seismic activity cannot account for both the
bursts and long-term torque changes unless the seismically active regions are
decoupled from one another. The idea is that since the observed changes in
spindown rate do not correlate with burst activity then the physical mechanisms
behind either phenomena are more likely unrelated. Whether the magnetar model
is able to cope with this observational requirement is not apparent.

\section{Concluding remarks}
\label{conclusions}

If a hypermagnetized neutron star could somehow be formed in a supernova
explosion, the abrupt amplification of its magnetic field will drive it into
collapse. Since magnetic flux is dissipated during the implosion via a process
analogous to the Sun's \textit{coronal mass ejections} (as in Malheiro et
al.~\cite{malheiro04}), the most likely outcome of such a collapse may give as
a remnant a strange star of canonical magnetic field, black hole or even an
exotic black string. Once again, the violent ``consumption" of the neutron star
nuclear material may trigger the emission of a gamma-ray burst of overall
energy $\sim10^{52}$~erg, inasmuch as in the quark-nova model (Ouyed et
al.~\cite{ouyed02}).

In the case of neutron stars such as the suggested magnetars, the anisotropy of
pressures must naturally develop, and the condition for the collapse in the
direction perpendicular to the dipole magnetic field could be satisfied
(Fig.~\ref{fig2}) for the typical values of density and magnetic field strength
routinely quoted for these stars.

To summarize, we have presented a consistent theory to discuss the stability of
compact remnant stars whose structure is dictated by a combination of quantum
and magnetic effects. We have shown that, in general,  the theory agrees quite
well with current observational data for magnetic WDs (Shapiro \&
Teukolsky~\cite{Shapiro}), canonical pulsars (Lorimer~\cite{lorimer99}), and
the realistic relativistic models of Cardall, Prakash \& Lattimer~\cite{CPL}. A
major outcome is the fact that some special configurations of highly magnetized
white dwarfs could collapse triggering explosions similar to a SNIa. Besides,
some neutron stars endowed with superstrong magnetic fields are shown to be
naturally unstable, and therefore should collapse, for their quoted surface
magnetic fields. This would drive a powerful supernova explosion followed by a
gamma-ray burst, too.

\begin{acknowledgements}
We are very grateful to Jo\~{a}o Pulido for stimulating discussions and helpful
comments. One of us (A.P.M.) would like to thank the ICTP, IAEA and UNESCO, and
CFIF (Lisbon, Portugal) for their financial support and hospitality. The work
of R.G.F. has been supported by \emph{Funda\c{c}\~{a}o para a Ci\^{e}ncia e a Tecnologia}
under the grant SFRH/BPD/1549/2000. H.J.M.C. is a fellow of the
\textit{Funda\c{c}\~{a}o de Amparo \`{a} Pesquisa do Estado do Rio de Janeiro} (FAPERJ),
Brazil, under the contract E-26/151.684/2002, and thanks OEA-ICTP for
hospitality in Havana through NET-35.
\end{acknowledgements}

\end{document}